# New quantum theory of the Vavilov-Cherenkov radiation and its analogues

S.G. Chefranov

A.M. Obukhov Institute of Atmospheric Physics RAS, Moscow, Russia;
schefranov@mail.ru

## Abstract

The general theory of non-bremsstrahlung radiation by non-equilibrium medium is proposed on the base of the new (PRL, 2004; JETP, 2004) quantum theory of Vavilov-Cherenkov radiation (VCR), where Abraham's form for the true photon momentum in the medium is used (only which allows the introduction in the energy-momentum balance equations of the effective real photon mass in the medium). This gives the possibility to consider the ejection of VCR quantum directly by medium, as complies with known assumptions of the microscopic mechanism of VCR (I.E. Tamm, 1939 and etc.). On the base of the Abraham theory, it is stated the new value of threshold velocity $V_{th} = c/n_*(n)$ of the charged particle, when only for $V > V_{th}$ the emission by medium of the VCR or its analogues is take place for $n>1$ and also for $n<1$ (c is light velocity in vacuum, n is the refraction index of isotropic medium in equilibrium state). It allows explaining of the observed "subthreshold" VCR effects and its analogues. It is show that for $n<1$, value $n_* > 1$ in the expression for $V_{th}$, that allows VCR mechanism realization when emitting transversal electromagnetic waves in plasma and also while emitting by a nuclear medium interacting with sufficiently fast $K^+$ and $K^0$ mesons. Propose the possibility for using of the new VCR theory in relation with super radiation in Bose-Einstein condensates and with the Compton effect in medium for $n<1$ and $n>1$. Near the same consideration on the base of the new VCR quantum theory is proposed for the modification of the Landau superfluid threshold velocity theory.

# New quantum theory of the Vavilov-Cherenkov radiation and its analogues

S.G. Chefranov

Content
1. Introduction
2. Quantum VCR and VCRA theory
3. Generalization of the NBR theory
4. Theory of super radiation in Bose-Einstein condensates
5. The theory of Compton effect in medium
6. The Landau threshold velocity of superfluid at finite temperatures
7. Conclusions

## 1. Introduction

It is known [1, 2] that mechanism of the Vavilov-Cherenkov radiation (VCR) effect corresponds to the some kind of non-bremsstrahlung radiation (NBR), since it is not related with necessity of the charged particle velocity V to change at all. According to the microscopic view [1, 2], the very non-equilibrium medium itself is the VCR emitter when V>$V_{th}$. However, in [1] and in consequent works (see [2, 3]) this concept did not get necessary development. It provided only in the new quantum VCR theory [4, 5] and in the new microscopic VCR theory [6].

Actually, in the quantum VCR theory [3], as in the macroscopic VCR theory [1] (where equilibrium stationary state of the medium is considered only), the same value for $V_{th}$ is obtained: $V_{th}=V_{th0}=c/n$, where c is light speed in vacuum, and n is the refraction index of the isotropic medium, being in the equilibrium state for n>1. In [3], however, VCR photon is emitted not by the medium, but by the particle itself. And in the energy-momentum balance equations, considered in [3] (see also given in [3] references), fully not counted any energy change ΔE of the medium related to emitting by it of the VCR quantum. This leads also to the distinction of the quantum VCR theory [3] from theories of other kinds of NBR [7-9]. Actually, due to the implicit assumption in [3] that ΔE=0, it is already not possible to use the "true" (see [10]) value of photon momentum in the form corresponding to Abraham's theory. Instead of it, in the momentum balance equation in [3], it is used the value of photons pseudo-momentum, that corresponds to the Minkowski theory. The latter theory actually more than 60 years was equal in rights alternative to the Abraham theory, which however already about 40 years ago has got direct experimental confirmation and public recognition [10]. In relation to this, in [11], for Minkowski's theory is stated the limit of its formal applicability only for the stationary equilibrium processes in the medium with electromagnetic fields participation. That is why, it is not surprising that conclusion of the quantum VCR theory [3] (based on the Minkowski theory) coincide with the conclusions of the macroscopic stationary VCR theory of Tamm-Frank [1].

Thus, the both VCR theories [1, 3] can't define threshold of non-stationary process of VCR emitting by the non-equilibrium (in local) medium, but correspond to the description of arising after VCR emitting equilibrium stationary state of the medium and VCR. Actually, in [4, 5] it is shown (see in [4, 5] Table of comparison with Cherenkov's experiment) that value $V_{th0}$

=c/n, obtained in [1, 3], well fits only to the position of interferential VCR maximum, but not to the VCR cone boundary, related to the VCR effect threshold.

In works [12-17], this resolution is not counted and threshold velocity $V_{th0}$ value is used for the description of observed VCR effects and its analogues in other media. However, it is found out that mentioned effects are observed already when $V < V_{th0}$, i.e. they have "sub threshold" character from the point of view of macroscopic VCR theory of Tamm-Frank [1] and quantum VCR theory [3].

To eliminate this discrepancy of observation data and theory [1, 3], in [12, 13, 17] for considering VCR analogue in the nuclear medium, it is introduced a concept of new mechanism of coherent particle emitting, dual coherent particle emission (DCPE). In [12, 13, 17], instead of $V_{th0}$, a new value is proposed: $V_{th}=V_{th0}=c/nn_1$, where $n_1$ is the refraction index of the wave field of a "charged" particle itself in the considered medium.

However in [12, 13, 17], as in the VCR quantum theory [3], the quantum of the VCR analogue (VCRA) is emitted by the particle itself. Accordingly, for the particle momentum and for VCRA quantum momentum in the medium with $n>1$ and $n_1>1$, in [11, 12, 16], it is used only Minkowski's representation without any counting of the pointed above applicability limits, established in [11] for Minkowski's theory itself.

In the result, in [12, 13, 17], for example, can't get explanation observed in nuclear medium threshold VCRA effect from $K^+$ and $K^0$ mesons, interacting with that medium for energies corresponding to condition with $n<1$.

In the present work, we propose development of the new quantum VCR and VCRA theories based on VCR theory [4, 5] and modification of theory [12, 13, 17], in which Abraham's representation now is used for the particle momentum. Meanwhile, observed "sub threshold" arising of VCR and VCRA effect, and also VCRA effect from $K^+$ and $K^0$ mesons now get explanation.

More over, "sub threshold" effects and VCRA from $K^+$ and $K^0$ mesons may get explanation also for $n_1=1$, i.e. already on the base of the value $V_{th}$, obtained in [4, 5] and fitting observation data (see Table in [4, 5]):

$$V_{th} = c/n_*, n_* = n + \sqrt{n^2 - 1}, \qquad (1)$$

where in (1) $n>1$ (for $n<1$ in (1) n must only be replaced by 1/n). Expression (1) is obtained in [4, 5] on the base of Abraham's form for photon momentum value in the medium with $n \neq 1$:

$$p = p_a = E/cn, \qquad (2)$$

where E is the photon energy in the medium with $n>1$ (for $n<1$, in (2), as in (1), it is only necessary to replace n by 1/n).

In the present work, we get generalization of (1) taking into account (2) and ideas of [12, 13, 17] about necessity of counting distinction from unity of refraction indexes for the wave fields of particles initiating VCR and VCRA, emitting by the medium. On the base of proposed quantum VCR (VCRA and NBR) theory it is also considered here the new theory of super radiation in the Bose-Einstein condensate (BEC), new theory of the Compton effect in medium and the new modification of the Landau superfluid threshold velocity.

## 2. Quantum VCR and VCRA theory

1. Let in laboratory frame of reference, in which uniform medium is not moving, laws of energy and momentum conservation for VCR and VCRA effects have the form [4, 5]:

$$E_1+\Delta E = E_0+E_2, \qquad (3)$$
$$\mathbf{P}_1=\mathbf{P}_0+\mathbf{P}_2, \qquad (4)$$

In (3), (4), values $(E_1, \mathbf{P}_1)$ and $(E_2, \mathbf{P}_2)$ correspond to energy and momentum of the particle before and after creation by the medium of VCR or VCRA quantum with energy $E_0$ and momentum $\mathbf{P}_0$. For example, for VCRA in the nuclear medium, state $(E_2, \mathbf{P}_2)$ may already correspond to some new particle, which not necessarily coinciding with the particle in the initial state with $(E_1, \mathbf{P}_1)$.

In the energy balance equation (3), as in [7,8] (see (1) in [7] and (4) in [8]), value $\Delta E$ is non zero. In (3), $\Delta E$ is related to the change of the medium energy caused by emitting from it of the VCR or VCRA quantum. In [3], inversely, it is assumed that $\Delta E$ in (3) is zero valued.

When defining values of momentum $p_\alpha$ ($\alpha=0,1,2$) in (4), we use contrary to [12,13,17], generalization of Abraham's representation (2). This generalization already takes into account suggested in [12,13,17] introduction of refraction indexes $n_\alpha$ ($\alpha=0,1,2$) for the wave fields of all particles participating in energy-momentum balance (3),(4), and has the form:

$$P_\alpha = \sqrt{E_\alpha^2 - m_\alpha^2 c^4}/cn_\alpha, \qquad (5)$$

where $n_\alpha > 1$ and $m_\alpha$ is the rest mass in vacuum of the particle with number $\alpha=0,1$ and 2 in the system (3), (4). When $n_\alpha < 1$ in (5) we must only replaced $n_\alpha$ on $1/n_\alpha$. For $m_0=0$, expression (5) for $\alpha=0$ exactly coincides with (2) for $n_0=n\neq 1$, when $E_o=E$ that is photon energy in the medium. Contrary to (5), in [12,13] for $P_\alpha$ in (3), (4) is used the Minkowski form $P_\alpha = n_\alpha\sqrt{E_\alpha^2 - m_\alpha^2 c^4}/c$, when $n_\alpha > 1$. More over in [12, 13] $\Delta E = 0$ in (3), (4).

Changes for the value of the medium momentum as a whole we neglect in (3), (4). Actually, mass and energy of the whole medium significantly exceeds masses $m_\alpha$ and energies $E_\alpha$ of the particles (with numbers $\alpha=0,1$ и 2) participating in VCR or VCRA. We assume negligible also corresponding to VCR or VCRA effect change of the medium momentum that in the result is not reflected in the energy-momentum balance equations (3), (4).

At the same time, as shown in [4, 5], it is very important to count in (3) the changes of medium energy $\Delta E$, caused by emitting by it of VCR or VCRA quantum.

In [4, 5], it is assumed estimate of the value $\Delta E$, based on conjecture that this value can't be less than energy corresponding to finite and real rest photon mass got by a photon in any medium with $n\neq 1$ according to (2).

For each particle counted in balance (3), (4) it is possible (as in [4, 5] on the base of (2)) to introduce, using (5), the following representation for the rest mass $M_\alpha$, which has a particle with number $\alpha$ (in the medium with $n_\alpha\neq 1$ for $\alpha=0, 1, 2$):

$$M_\alpha = \sqrt{E_\alpha^2 - P_\alpha^2 c^2}/c^2, \qquad (6)$$

where $P_\alpha$ is from (5) for $n_\alpha > 1$, and in the case of $n_\alpha < 1$, it is necessary only to replace in (6) $n_\alpha$ by $1/n_\alpha$.

Using (6) it is possible by analogy with approach of works [4, 5] to define an estimate of the lower boundary $\Delta E_{min}$ for the value $\Delta E$ used in the energy balance equation (3):

$$\Delta E \geq \Delta E_{min} = c^2 (M_0+M_2-M_1), \qquad (7)$$

where $M\alpha$ ($\alpha$=0, 1, 2) are defined from (5), (6). The estimate $\Delta E_{min}$ from (7) corresponds to such necessary change of the medium energy that allows possibility of emitting by the medium of particles with masses $M_0$ and $M_2$ when initially a particle with mass $M_1$ was absorbed by medium.

In particular, as in [4, 5], when emitting by the medium of a Cherenkov photon with mass $M_0 = E_0\sqrt{n^2-1}/nc^2$ (with its zero vacuum rest mass $m_0$=0 and $n_0$=n>1; for n<1 we only replace n on the 1/n in this expression )in (7) the value $\Delta E_{min} = M_0 c^2$, if in (6) and (7) for electron with $n_1$=$n_2$=1 we have $m_1$=$m_2$=m, where m is the rest mass of electron in vacuum.

More strong definition of the value $\Delta E$ (and effective mass $M_0$) in the energy conservation law (3) is made in [6] on the base of microscopic theory of non-equilibrium processes in the medium, providing emitting of VCR quantum by the medium.

2. From (3) - (5) under condition of meeting equality $\Delta E = \Delta E_{min}$ in (3) (where $\Delta E_{min}$ is defined from (6), (7)), it is possible for $n_\alpha \geq 1$ ($\alpha$=0, 1, 2) to get the following expression for cosine of angle between momentum vectors $\mathbf{P}_1$ and , $\mathbf{P}_0$:

$$\cos\theta_{10} = \frac{n_0 n_1 (2E_0 E_1 y/n_2^2 + (1/n_0^2 - y^2/n_2^2)E_0^2 + (1/n_1^2 - 1/n_2^2)E_1^2 + c^4(m_2^2/n_2^2 - m_1^2/n_1^2 - m_0^2/n_0^2))}{2\sqrt{(E_0^2 - m_0^2 c^4)(E_1^2 - m_1^2 c^4)}},$$

$$\Delta E_{min} = B \pm \sqrt{B^2 + (E_1-E_0)^2(n_2^2-1) + m_2^2 c^4 - n_2^2(A_1-A_0)^2};$$

$$B = (E_1-E_0)(n_2^2-1) - n_2^2(A_1-A_0); A_\alpha = \sqrt{E_\alpha^2(n_\alpha^2-1) + m_\alpha^2 c^4}/n_\alpha; \alpha=1,0 \qquad (8)$$

where y =1-$\Delta E_{min}/E_0$. For the case $n_\alpha$, in (8) it is necessary only to replace $n_\alpha$ by $1/n_\alpha$. In particular, for $n_1$=$n_2$=1, $m_1$=$m_2$=m and $m_0$=0 formula (8) coincides with formula (6) of the work [5] for $n_0$=n>1. For this coincidence may take place in (8) we must take into consideration only the case with sign plus, when we may reduce (8) to the results of [4, 5].

While modifying formula (8), with replacement of $n_\alpha$ by $1/n_\alpha$ is performed for all $\alpha$=0, 1, 2, it for y=1 formaly coincides with formula obtained in [12]. In [12], meanwhile, however, vice versa, it is considered that only $n_\alpha > 1$ are take place. It is the result of well known coincidence for the Abraham and Minkowski forms of photon momentum for the cases with n<1[10].

In the limit $E_0 \ll E_1$ and for $n_2$=$n_1$, $m_2$=$m_1$ and $m_0$=0 for any $n_0$=n≠1 and $n_1$≠1 instead of (8), we get(for n>1 and $n_1 > 1$):

$$\cos\theta_{10} = cynn_1/V_1, \qquad (9)$$

where $V_1$ is velocity of the particle with initial energy $E_1$ in (3). For cases with n<1, in (9) it must be replaced n on 1/n and so on for the cases with $n_1 < 1$.

As it was pointed above, for y=1 in the case of n<1, $n_1$<1 the modification of formula (9) formal coincides with the one got in [12], where it however is used only for cases with n>1 and $n_1$>1. Pointed out formula is considered in [12] in relation with interpretation of observation data of "sub threshold" VCRA effect realization. And only for $n_1$>1 (and n>1), in [12], it is succeeded to establish agreement of the theory with the data of corresponding experiments on VCRA in nuclear medium.

Contrary to [12], on the base of representation (9), considered even for n>1 and $n_1$>1, it is possible to achieve the same coincidence with pointed out experimental data, but now only due to the choice of sufficiently small value y<1 in (9). So, for example for noted in [12] values $n_1$= 1,00125 and $n_0$ = n = 1,00113 (for pions causing VCRA when $P_1$ = 1 Gev/s), formula (9) gives the same fitting with experimental data as theory [12] already for y=0,995, i.e. for $\Delta E_{min}/E$=0,005. In the general case, formula in [12] and (9) for $n_1$>1 and n>1 lead to the same (just formally) expressions if we take in (9) the formal relation: $y = 1/n^2 n_1^2$.

From (9) and condition $|\cos\theta_{10}| \leq 1$, we get the following new representation for the threshold velocity of a particle initiating VCR or VCRA (for n>1, $n_1$>1; for $m_1=m_2$, $n_1=n_2$ and $(E_1 - E_2)/E_1 \ll 1$):

$$V_1 > V_{th} = cn_1 ny = cn_1 n(1 - \sqrt{1 - 1/n^2(1 - m_0^2 c^4/E_0^2)} + ((E_1 - E_2)/E_1)\sqrt{1 - 1/n_1^2(1 - m_1^2 c^4/E_1^2)})$$
(10)

In particular, when $E_2 \to E_1$ and $m_0$=0, from (10), we get $V_{th} = cn_1/n_*$, where $n_* = n + \sqrt{n^2 - 1}$ for n>1. This expression for the case of $n_1$=1 exactly coincides with the one obtained in [4, 5] for n>1. For n<1, it is necessary only to replace in (10) n by 1/n, and for $n_1$<1, it is necessary similarly to replace in (10) $n_1$ by $1/n_1$.

In (10) value $n_1 ny$<1 and corresponding threshold condition for $E_1$ has the following form:

$$E_1 > E_{1th} = m_1 c^2 / \sqrt{1 - y^2 n^2 n_1^2}, \text{ for n>1 and } n_1>1, y<1/nn_1.$$

For n<1 or $n_1$<1 here it is necessary, as in (10), to replace n by 1/n and $n_1$ by $1/n_1$. In this relation, developed above theory has more broad range of applicability than VCR theory [3] and its generalization [12,13], which exclude from consideration effects of VCRA for n<1 or $n_1$<1.

In the result, contrary to [12,13], on the base of the suggested above theory of VCR and VCRA, it is possible the new consideration of the effect of emitting by the medium a quantum of VCRA caused by interaction of the medium with sufficiently fast $K^+$ and $K^0$ mesons. Actually, for the observed radiation in the entire spectrum range of energy, the refraction index $n_0$<1. In [12, 13], emitting of $K^+$ and $K^0$ mesons therefore is considered as not solved yet problem because of impossibility of application of theory [3] in the cases with $n_0$<1.

3. Let us note also possibility of use of (9), (10) also for interpretation of observed «sub threshold » VCR effects in dielectrics (see section A in [15]). Meanwhile, exceeding $\Delta\theta$= $\theta_{max}$ - $\theta_s$ =$2,1^0$ of the limit angle of the observed VCR cone ($\theta_{max} \approx 36,9^0$) over the theoretical value ($\theta_s \approx 34,8^0$ according to VCR theory [3]) corresponds to the estimate $yn^2$=0,974 of the presented theory.

Actually, according to (9), for $n_1$=1, we have cos $\theta_{max}$ =cny/$V_1$ and $\cos\theta_{max}/\cos\theta_s = n^2 y$ (where for theory [3], $\cos\theta_S$ =c/$nV_1$). For n=1,505 (for wave length $\lambda$=4*$10^{-5}$ sm. [15]), we have the formal empiric estimate y=0,43 or $\Delta E/E_0 \approx 0,57$, which is rather close to the theoretical estimate $\Delta E_{min}/E_0 = \sqrt{n^2 - 1}/n \approx 0,75$, obtained for the n=1,505 from (7). From the other side, estimate of the value $\Delta\theta$ on the base of suggested in [15] theory (explaining observed «sub thresholde» VCR demonstration by effects of bremsstrahlung radiation from the end parts of trajectory, see also [1]) several times exceeds observed quantitative value $\Delta\theta$ (see (5.1) in [15] for $\varepsilon$=0,275, $\beta_n$=1,217 and $1/\gamma_n\beta_n = 0,57$).

Thus far, in the frame of the new quantum VCR and VCRA theory being developed in the present work, it is found out possibility of explaining of observed «before threshold » VCR and VCRA effects without use of any auxiliary ideas going out of consideration of the very non-bremsstrahlung mechanism of emitting of VCR and VCRA quanta by non-equilibrium medium.

In this new VCR and VCRA theory the important meaning have the possibility to introduction on the base of the Abraham theory of non zero real value for effective photon mass in medium which appears in parameter y, arising in (8)-(10) . Instead it, in [18] the VCR theory provided on the base of using the introduction (in connection with the Proca theory) of non zero real photon mass in vacuum without any relation to the Abraham theory. The corresponding to (8)-(10) formulas in theory [18] are thus very different from (8)-(10) .

## 3. Generalization of NBR theories

1. Laws of energy and momentum conservation (3), (4) for VCR or VCRA, where $\Delta E \neq 0$ coincide, as already noted, also with ones used in the theory other kinds of non-bremsstrahlung radiation (see [7, 8]). Value $\Delta E$, however, may get different physical sense, as e.g., for the cases of the combine [7] and different kinds of the transitory [8, 9] radiation. We exclude from consideration here only the case of the transitory radiation emerging due to the transition of charge particle from one medium to another, when actually is possible also effect of particle acceleration and it is not possible to speak about pure non-bremsstrahlung character of radiation by the medium itself.

Contrary to the developed in § 2 theory, in [9] value $\Delta E$ is already defined via electromagnetic energy change in the medium $\Delta W_0^q$, caused by arbitrary small jump of dielectric permeability $\Delta \varepsilon(t)$ at some moment for spatially uniform medium. However, used in (7) estimate from below $\Delta E_{min}$ may be related with value by the following relationship (in assumption that in (3), $\Delta E = \Delta E_{min}$ for $\Delta E_{min}$ from (7) for $M_1 = M_2$ and $n_0 = n < 1$, when $\Delta E_{min} = \hbar \omega_p$ [3,4], where $\omega_p$ is the plasma frequency):

$$\Delta E_{min} = 2 |\Delta W_0^q| / \alpha \gamma, \qquad (11)$$

where $\gamma = 1/\sqrt{1-V^2/c^2}$, $\alpha = e^2/\hbar c$ is a constant of the thin structure. Hence, from (11), we get that only under condition $\gamma = 2/\alpha$, the exact coincidence of $|\Delta W_0^q|$ and $\Delta E_{min}$ is takes place.

In [9], value $\Delta W_0^q$ does not depend on specifics of jump $\Delta \varepsilon(t)$ and is related with plasma frequency $\omega_p$ by the very relationship (11). That is why, it is possible also, vice versa, to consider as allowed such VCR mechanism that some change $\Delta \varepsilon(t)$ mutually related with creation of VCR quantum by non-equilibrium medium got out from equilibrium state (where $\varepsilon = n^2 = $const for isotropic medium being in equilibrium state) by the non stationary electric field of a sufficiently fast charged particle [6]. It also complies with considered in [8] mechanism of radiation emitting and with suggested in [4, 5] (for n<1) VCR mechanism in plasma as conversion of condensate (extremely long-wave) plasmon with energy $\hbar \omega_p$ in transversal Cherenkov,s photon.

For non-zero value $\Delta E$ in (3), as it takes place in different theories of non-bremsstrahlung radiation, it is formally acceptable to use representation of momentum in the medium both in Minkwski's and Abraham's forms. However, as in the VCR theory [3] (where $\Delta E=0$ and, as noted, the using of Abraham's form leads to the results which have not any physical sense), in the pointed different variants of non-bremsstrahlung radiation theory nevertheless traditionally the very Minkowski's representation is used for photon momentum in the medium. This, as already noted, looks not reasonable according to Minkowski's theory applicability limits stated in [11]. That is why, used in [4,5] and in the present work Abraham's representation for photon momentum (and momentum of other particles) in the medium is the only acceptable when defining threshold of non-equilibrium processes in the medium leading to emitting by medium of non-bremsstrahlung radiation of any type. Quantum theory given in §2, actually describes all kinds of non-bremsstrahlung radiation by non-equilibrium medium.

## 4. Theory of super radiation in Bose-Einstein condensates

On the base of developed above and in [4, 5] new approach to VCR theory it is giving description of energy-momentum conservation laws for practically all possible kinds of non-bremsstrahlung radiation. Let us consider also this theory as a variant of possible explanation for observed threshold process of super-radiation in Bose-Einstein condensates (BEC) [19-22].

This super radiation in BEC is observed under the laser irradiation action on BEC by the pumping photon with energy $E_{ph}=\hbar\omega_{ph}$. In the existing theory of super radiation in BEC (see, e.g., [22]) still absent any explanation of threshold mechanism of this phenomenon related with observed accelerated motion of the part of atoms of BEC.

According to [4,5] and (9), (10), when realizing VCR (or VCRA) effect, actually, it is possible simultaneous with VCR (or VCRA) quantum emitting the initiation of acceleration motion for the charged particle, interacting with the medium during VCR (or VCRA). It takes place when the value of parameter y in (9) becomes negative. Similarly, it is possible to suppose that super radiation of a photon with energy $E_0=\hbar\omega_0$, observed simultaneously with acceleration of motion of atoms of condensate is performed in accordance with considered in § 2 mechanism of VCR and VCRA for y<0, when $E_{ph}= \Delta E > E_0$, i.e. for $\omega_{ph}>\omega_0$.

Thus according to (9), (10) necessary condition of realization of super radiation BEC is the very inequality $\omega_{ph}>\omega_0$, which on the base of [19] may be represented in the form:

$$R > R_{th} = 8\pi P / 3\hbar\omega_0 \sin^2\theta_j N_0 N_j Q_j . \qquad (12)$$

In (12), R is the speed of Raley scattering for BEC atom which is proportional to intensity of laser pump beam, $\theta_j$ is the angle between directions of spreading of super radiation and polarization of the pumping beam, $Q_j$ is spatial angle of super radiation with power P, $N_0$ is number of all BEC atoms, $N_j$ is the BEC atom number getting acceleration under super radiation.

If in (12) to take (see [20,21]) values $P \cong 0,8*10^{-3}$ Vatt/sm$^2$ and $N_0 \cong 10^6$, $Q_j \cong 1,9*10^{-4}$, then from (12), we get $R_{th} \cong 10^{22}/N_j\omega_0$. For example, for $N_j = 10^4$ and $\omega_0 = 10^{15}$ sec$^{-1}$ from (12), it follows an estimate $R_{th} \cong 10^3$ sec$^{-1}$, that complies well with the observed threshold value R (see Fig.3 in [19], where $R_{th}^{exp.}=0,7*10^3$ sec$^{-1}$). In [22], where it is noted absence of explanation of threshold on energy super radiation by BEC atoms, it is pointed on the presence of relationship $\omega_0 = c|k| - \omega_{ph}$ for irradiated mode with frequency $\omega_0$ and wave vector k in the system rotating with pump frequency $\omega_{ph}$. From the inequality $\omega_0 < \omega_{ph}$ (i.e. criterion of super radiation of VCR+ type), it follows not only condition (12), but also additional restriction on the value $\omega_{ph}$ from below because it must also hold inequality $\omega_{ph} > ck/2$.

Let us note, that under more strict condition $\omega_{ph} > ck$ value of energy $E_0=\hbar\omega_0$ becomes negative and creation of photons of super radiation becomes energetically beneficial as also for other systems with realization of mechanism with dissipation instability in the presence of negative energy [4, 5, 23].

Observed anisotropy of BEC super radiation may be also explained in the frame of the new VCR theory on the base of (9), (10).

Of course, suggested here analogy of mechanisms of super radiation and VCR needs further refinement. It must take into account specifics of the process of energy change ΔE. Here, it is assumed that the medium energy change that leads to emitting of a quantum of super radiation with necessity must exactly be compensated by coming from outside energy of quantum $\hbar\omega_{ph}$ of the laser pump beam that provides introduced above definition of the value of ΔE via pump photon energy $\hbar\omega_{ph}$.

Thus, contrary to adopted now point of view, BEC atoms may be accelerated not due to direct acting on them by the pump photon, but as the very consequence of simultaneous realization of VCR+ effect by the medium for y<0 in (9). Super radiation initiated in the medium by accelerated BEC atoms is the consequence of the medium instability (with respect to VCR effect realization), caused by its non-equilibrium state, formed under action of an external (laser) pump photon.

## 5. The theory of Compton effect in medium

On the base of the energy-momentum balance equations (3), (4) it is also possible to provide new theory of the Compton effect in medium with $n_\alpha \neq 1$, as it makes above for super radiation in BEC (and in close analogy with BEC). Lets now in (3), (4) the values which having index $\alpha = 1,2$ will refers to the photon in its initial and final states, correspondent. The value 0 for this index now is refers to the electron of medium. As the result, it may be obtained the same formula as (8), where now we have dependence for the cosine of angle between the initial and final directions of photon momentum. In this version of (8) the values of the photon mass in rest $m_2 = m_1 = 0$ and $m_0 = m$, where m- mass of electron in rest. Our theory is formulated for the medium without dispersion. When it is necessary to take into account the dispersion we must only replace in all formulas: $n_\alpha \to n_\alpha + E_\alpha \frac{\partial n_\alpha}{\partial E_\alpha}$.

Lets consider, for simplicity the case, when $n_0 = 1, M_0 = m; n_1 = n_2 = n$ in this new version of (8), which now have another form (when in (3),(4) now $E_1, E_2$ -is energy of photon before and after Compton effect and electron obtained energy $E_0 = mc^2 / \sqrt{1 - \frac{V^2}{c^2}}$ and module of momentum $|P_0| = mV / \sqrt{1 - \frac{V^2}{c^2}}$ for $\Delta E_{\min} = (E_2 - E_1)\sqrt{n^2 - 1}/n + mc^2, n > 1$ from (7)):

$$x^2(1 - N^2) - 2x(1 - \cos\theta + nN/\varepsilon_1) + 2(1 - \cos\theta) = 0,$$

$$where N = (n - \sqrt{n^2 - 1}), \varepsilon_1 = E_1/mc^2, x = (E_1 - E_2)/E_1$$ (13)

When N=n=1, the equation (13) have solution, which give relativistic generalization for known Compton effect theory:

$$x = 1 - \frac{1}{1 + \varepsilon_1(1 - \cos\theta)}$$ (13.1)

Only in the limit $\varepsilon_1 \ll 1$ from (13.1) we may obtain the well known formula: $x = \varepsilon_1(1 - \cos\theta) + O(\varepsilon_1^2)$. If we consider only the case with $\Delta E = \Delta E_{\min} = mc^2$ in (3),(4) (in limit of very small effective mass of photon in Compton effect in medium), it is necessary to replace in (13) N on n. In other cases we have from (13) the common solution:

$$x = x_\pm = (D \pm \sqrt{D^2 - 2\varepsilon_1^2(1 - N^2)(1 - \cos\theta)})/\varepsilon_1(1 - N^2); D = nN + \varepsilon_1(1 - \cos\theta)$$ (13.2)

In the limit $\varepsilon_1 \ll 1$ from (13.2) we have the more simple modification of Compton effect theory in medium with n>1:

$$x_- = \varepsilon_1(1 - \cos\theta)/nN + O(\varepsilon_1^2); x_+ = 2nN/\varepsilon_1(1 - N^2) + 2(1 - \cos\theta)/(1 - N^2) + O(\varepsilon_1),$$ (14)

where only for $x_-$ in the case with n=1 formula (14) gives the well known presentation of the Compton effect theory.

The unusual case when electron is not exist in medium before the realization of such "Compton effect" with the new electron arising (when for n>1 in (3) $\Delta E = (E_2 - E_1)\sqrt{n^2 - 1}/n$), is more natural to tied with the new quantum VCR type theory for the beta radioactivity effect in the another paper.

For the case of medium with n<1 in all formulas the value n must only be replaced by 1/n. Thus for both cases n>1 and n<1 we obtain that $x_-$ is increased when n is near unit and changes up (for n>1) or down (if n<1) from unit in small degree. In the case n>1 this tendency in dependence on n in (14) is take place in contrast to the result of [24], where instead of (14) it is obtained( from the another type base theory) the next formula (see formula (70) in [24]): $x = \varepsilon_1(1 - n^2 \cos\theta)$, n>1.

The effect of medium's refractive index on the Compton effect may be tested in experiment by measuring directly the resulting photon energy difference or its frequency changing. In the last case in (13.2) and (14), according to [25], for n>1 it must be taken the next presentation of photons energies: $E_1 = n^2 \hbar\omega_1, E_2 = n^2 \hbar\omega_2$, where $\omega_1, \omega_2$ - corresponding frequencies.

In future it is also interesting to consider the conditions for the cases with negative values for x in the Compton type effect in medium. In [24] for n>1 this negative values of x are obtained only for some angles ( when $1 - n^2 \cos\theta < 0$ in formula (70) of [24]) and this anomalous Compton effect also proposed in [24].

It may be also note the analogy of the considered above new theory of super radiation in BEC and presented here new theory of the Compton effect in medium with n>1 and n<1.

## 6. The Landau threshold velocity of superfluid at finite temperatures

On the base of the quantum VCR theory, formulated by the resolving of the energy-momentum balance equations (3), (4), it is also possible to give the new generalization of the Landau theory for threshold velocity of the superfluid. More over the new VCR theory in [4, 5] was formulated by us first of all as the relativistic generalization of this Landau's theory [26, 27], where we in additional also take into account the important role of the medium energy changing $\Delta E$ in (3) during VCR effect.

Lets now consider (3),(4),where now the values $E_1 = m_k V_s^2/2, E_2 = m_k V_{1s}^2/2$ are corresponding to the initial and final kinetic energy of the capillary walls with macroscopic mass $m_k$ (in the reference frame, where superfluid velocity is zero and the capillary walls are moving with velocity $V_s$ )before and after threshold arising in super fluid of elementary excitation with energy $E_0(P_0)$, where $P_0$ is the momentum of this excitation. Now the value $\Delta E = \mu(t)V_s^2/2$ in (3) is the energy difference of superfluid

13medium during this process of arising excitation at some $V_s > V_{th}$, when the state of superfluidity become unstable due to the attaching to the capillary wall of some small mass $\mu(T)$ of superfluid at first stage.

Thus, in the limit $P_0 \ll m_k V_s$ it may be obtained from (3), (4), instead of (8):

$$\cos\theta_{10} = \frac{\Delta E - E_0(P_0)}{V_s P_0}, \qquad (15)$$

which giving the new generalization of the Landau criteria in the form:

$$V_s > V_{th} = \min\frac{|E(P_0) - \Delta E|}{P_0} \qquad (16)$$

Now in (15) the value $\theta_{10}$ is the angle at what vortex excitation inclined in the flow of superfluid from the rigid capillary wall or the wall of some other object moving in the frame of reference, where the velocity of superfluid is zero. In (16) the minimum must be taken over the values of the excitation momentum $P_0$, as in the Landau theory, formulated for zero absolute temperature T=0 [26, 27]. The important difference of (16) from the result of Landau theory [26, 27] is stated in the presence of non zero value $\Delta E$ in (16). This value may be actually significant, especially for non zero temperatures T>0. At T=0 it may have tendency to zero, when (16) exactly coincided for $\Delta E = 0$ with the Landau theory [26, 27].

In [26, 27] for $E_0(P_0)$ the energy spectrum of roton excitations is used and the value of $V_{th}$ is obtained, which is equal near 60m/s. This estimation from the Landau theory have large difference from the observed values 0.01-10m/s for T>0 and thus constitute the well known unresolved up to now problem of the superfluid threshold velocity determination [28, 29].

Here we give the new example of the derivation for this famous Landau's theory and also for it finite temperature modification, which is making on the base of the energy and momentum conservation laws only. In the known theories of superfluidity at T=0 and also at T>0 (see, for example, [30, 31]) these conservation laws do not considered at all. More over, in [30] writing about the Landau theory [26, 27] (where only Galilelian transformation is used ): "Landau pointed out that the creation of elementary excitations at zero temperature by interactions with walls (or moving object) violates energy and momentum conservation unless the velocity exceeds a certain well defined velocity $V_L$".

Thus, obtaining of (16) on the base of (3), (4) shows another view on the theory of superfluid threshold velocity for T=0 and also for T>0. The difference between the value of the $V_{th}$, obtaining from (16), and the value $V_L$ of the theory [26, 27] may be very large, when the value of $\Delta E$ in (16) become as large as the excitation energy $E_0$ in (16). This is give the possibility to state more good correspondence between (16) and experimental date, mentioned above, than it take place for the Landau theory [26, 27].

For example, lets consider the new superfluidity criteria (16) for elementary excitations of the roton type [27], when in (16) $E_0(P_0, T) = \Delta(T) + (P_0 - p_0(T))^2 / 2m_*(T)$, where $\Delta$-is the roton energy gap and $p_0$ is the momentum at the roton energy minimum, $m_* = 0,16 m_{^4He}$, $m_{^4He}$-is the mass of helium atom. From (16) in this case it may be obtained:

$$V_{th} = \frac{p_0 - \sqrt{p_0^2 - 2\Delta(\mu - m_*)}}{\mu - m_*}, \qquad (17)$$

which is coincide exactly with the result of the Landau theory [27], when $\mu = 0$ and in the limit $m_* \Delta \ll p_0^2$ from (17) it may be obtained well known Landau's threshold velocity $V_{th} = V_L = \Delta / p_0$. From other side in (17) we may consider the limit $\mu \gg m_*$, when $p_0 \approx \sqrt{2\mu\Delta}$ and $V_{th} = \sqrt{2\Delta(T)/\mu(T)} = V_{L0}(T)$, where at T=0 this value exactly coincided with the first Landau's estimation in [26] of the superfluid threshold velocity. In [26] value $V_{L0}(T=0)$ is obtained on the base of the using of energy spectrum of roton in the form $E_0(P) = \Delta + P^2/2\mu_0$, where $\mu_0 = \mu(T=0) \approx 7 m_{^4He}$.

It may be note, that not for all more large then (17) superfluid velocities the vorton type of excitations can arising from the boundary of superfluid. Indeed, from (16) for the roton energy spectrum we have the possibility for roton arising only when $V_{max} > V_s > V_{th}, V_{max} = \frac{p_0 + \sqrt{p_0^2 - 2(\mu - m_*)\Delta}}{\mu - m_*}$, where $V_{th}$ is from (17). When $V_s > V_{max}$ it is possible to arising of another types of excitations, but not for the roton excitations.

In this connection, lets consider, for example, the conditions for arising of the phonon-maxon type of excitations, when in (16) the energy spectrum of excitations have the form [32, 33]: $E_0(P) = E_{ph-max} = uP + \alpha P^3$, where u is the sound velocity and $\alpha = -u/3p_{max}^2 < 0$. Negative value of $\alpha$ is corresponding here to the stability condition of energy spectrum( see [33]) and $p_{max} \approx p_0/1,6$ is the value of the excitation momentum at the "maxon" point of energy spectrum [32]. From (16) now we have the next condition for the phonon-maxon excitations arising:

$$V_s > V_{th(ph-max)} = u(1 - P_*^2/p_{max}^2); x = P_*/p_{max}, (1+x^2)^2 = 4p_{max} x^3/3\mu u, \qquad (18)$$

where only for $\mu = 0$ the solution of equation for x in (18) gives x=0 and the well known result of the Landau theory[26, 27] in the case of pure phonon excitations. In the limit x≪1 also for the phonon excitations we have from (18) $x \approx x_0 = (3\mu u/4 p_{max})^{1/3}, V_{th(ph)} \approx u(1 - x_0^2)$. From the other side, for the maxon type excitations the condition (18) giving: $V_{th(max on)} \approx 2u\delta; x = 1 - \delta; \delta \approx \frac{1 - 3\mu u/p_{max}}{3(1 - 2\mu u/p_{max})} \ll 1$.

More detail evaluation of $\mu(T)$ and the corresponding new threshold superfluid velocity (17) and (18) will be done in the special paper. Now we may only note about the possible correspondence between the value $\mu(T)$ and the measured value of the thermal vortex activation over a potential-energy barrier (which is need for the creation of vortices in superfluid helium; see measurements [34] of the rate at which negative ions create vortex rings at 50<T<500mK with this energy barrier height $\Delta E_{bar} \cong 3,7 \times 10^{-16} erg$ ). Another possibility, with much more larger estimated value of $\mu(T)$, may be done on the base of measured energy $\Delta E_{phase-slip} = 1,2 \times 10^{-17} J$, which dissipated per isolated phase-slip event [29,35,36]. If, for example, we also, as for energy barrier $\Delta E_{bar}$, assume that $\mu V_{th}^2 / 2 \approx \Delta E_{phase-slip}$ it may be also obtain the empirical estimation of $\mu(T)$ for T>0.

Thus, we propose new modification of the Landau threshold superfluid velocity, which may used, as for small orifices (or short capillary) and fast moving ions in superfluid medium(when it is the most important of the temperature dependence for the relative large threshold velocity value), as for usually used long capillary(when the value of threshold velocity is much smaller and is main used of the Feynman critical velocity estimation). In the last case we may obtain the lower critical velocity only due to the increasing of the effective mass for spiral vortex structures , which are more energetically preference to arise in long capillary[37-39].

## 7. Conclusions

1. Thus far, as in [4,5], in the present work, new VCR and VCRA quantum theory gets development. It is taken into account that only Abraham's representation of the photon momentum in the medium may define actual value of effective photon mass obtained by it in any medium with $n \neq 1$.

Actually, for example, in [40], it is not taken into account when in relation with the fundamental problem of the photon localization in the medium, it is also introduced its effective mass $m_{eff} \neq 0$, obtained by the photon in the medium. For example, in the medium without dispersion (when the values of group and phase velocities are coincide, i.e. $V = V_{ph} = c/n$ in [40]), obtained in [40] for n>1 value $m_{eff}$ has the form

$$m_{eff} = \hbar \omega n \sqrt{n^2 - 1}/c^2, \tag{19}$$

which coincided with value $M_0$ in (6), when photon energy in the medium has the form $E_0 = \hbar \omega n^2$, proposed also in [25].

In [40], actually, when introducing (19), it is stated that if $E_v = \hbar \omega$ is the photon energy in vacuum then value $E_0 = E_v n^2$ defines photon energy in the medium with refraction factor n>1. However, permissibility of the expression $E_0 = E_v n^2$ is mistakenly related in [40] with very Minkowskiy's representation $p_m = E_0 n/c$ for the photon momentum in the medium with n>1 . In [25] proposed that for $p_m$, contrary to the pointed above expression $E_0$, only equation $E_0 = E_v$ takes place and energy of photon is not depending on n .

Actually, it is found out (see [25] and given there references) that expression of the form $E_0 = E_v n^2$ for the photon energy in the medium with n>1 is permissible only for Abraham's representation (2) for the photon momentum in the medium. At the same time, for the photon momentum in the medium in Minkowskiy's representation with necessity (see [25]) it must hold very equality of energies $E_0=E_v$. According to (6) (for a photon with $m_o=0$) for $P_0 = p_m$, instead of (19) one already gets as known [5], a complex, not real value $m_{eff} \neq M_0$ for n>1.

Thus far, only when using Abraham's representation (2), from (6) when $m_0=0$, we get finite and real value of the effective photon mass (see [4,5] and (6), (7)) in the medium with n≠1. It is important for the developed in [4,5] and in the present work VCR and VCRA theory generalizing theory of all kinds of non- bremsstrahlung radiation and also for resolution of the noted in [40] problem of the photon localization in quantum optics and mathematical physics.

2. From the other side, in the theory of combinational [7] and transitory radiation in non-stationary medium [8], when defining in (3) the value ΔE, it is not used an idea about effective photon mass in explicit form. It is related with that in [7, 8] (as also in [3, 9]), it is considered only Minkowskiy's representation for the photon momentum in the medium, for which value of effective photon mass is complex and its counting in (3) has not physical sense. However, result obtained for example in [7] (see (2) in [7]) for the value $\cos\theta_{10}$ (i.e. cosine of the angle of outgoing of quanta) has just the same dependency on ΔE, as also in (8) (for $n_1=n_2=1$) and in similar formulas [4, 5]. There is only significant difference from [7] in the form of dependency of $\cos\theta_{10}$ and accordingly Vth on the value n. It takes place due to that in [7], it is used Mikowskiy's representation, not Abraham's, for the photon momentum in the medium. Value of refraction factor n in [7], as in the present work is related to equilibrium medium not having excited atoms, very presence of which leads to ΔE≠0 in [7].

Thus, it is possible to consider investigated in [7, 8] radiation as kinds of VCR having common with VCR non-bremsstrahlung mechanism of radiation. Let us note, that in [7], as in (3), (4), it is considered only the case of spatially uniform medium (contrary to [8]) neglecting the impact of random non-uniformity. In the result, in [7], as in (4), it is not counted change of medium momentum value and there is significant similarity of the mechanism pointed in [7] with pointed in [4, 5] and in the present work VCR mechanism when "virtual photon is converted to the real one" (see [7]) for $V>V_{th}$ where $V_{th}$ is from (10).

There is also some similarity between considered in [7] and in [41] non-equilibrium states of medium interacting with moving particle. In [41], however, this non-equilibrium state of medium (with negative temperature) is defined not depending on the particle velocity contrary to [7] and suggested in [4, 5] and in the present work consideration.

In [8], value ΔE in (3) (in [8] see equation (4)) is related now, contrary to [7], with change frequency of the electromagnetic wave traversing the medium. In the result of traversing the medium by such a wave, properties of the medium can change in time, for example its dielectric constant (experiencing periodical displacements from its equilibrium value). In VCR mechanism, the role of such a wave is played by the particle electric field moving relatively to the medium and causing, essentially considered in [6, 7], excitation of the medium atoms. Frequency of such a wave $\omega_{ph}$ defines in particular energy $\Delta E=\hbar\omega_{ph}$, as in analogy the laser pump wave which under condition (12) provides super radiation of BEC atoms.

Thus, developed in [4, 5] and in the present work common approach to description of all NBR types on the base Abraham's theory allows with the help of balance equations (3), (4) to define threshold of corresponding radiation by non-equilibrium medium independent from details of microscopic mechanism of such radiation.

3. When refraction index is close to one, i.e. when n-1<<1, it seems from the first looking, that distinctions in conclusions about VCR threshold obtained from VCR theory [3] and

17from [4, 5] and (10) must be eliminated. However, as it is shown in [42], it is far not so. Actually, in theory [42] (based on [4, 5]), it is show that contrary to [43, 44] (using VCR theory [3]), it is possible realization in the modern epoch of VCR effect from photon gas of cosmic radiation interacting with relativistic particles of cosmic rays. This conclusion according to estimates [42], gives base also to explain observed cutting in the spectrum of cosmic rays energy being alternative to the known GZK cutting mechanism [45, 46].

4. More significant distinction of the new quantum VCR and VCR+ theory from the known VCR theory [1, 3], than distinction of values $V_{th}=c/n_*$ and $V_{th}=c/n$ for n>1, is obtained in [4, 5] and in the present work conclusion about possibility of realization of very VCR mechanism even in the medium with n<1. For example, efficiency of direct radiation of transversal electromagnetic waves in plasma due to the new counting of VCR mechanism by the medium is found to be significantly greater (see [5]) than known mechanism of transformation of longitudinal waves into transversal investigated e.g. in [47]. In the new VCR theory, there is transformation also but now of virtual longitudinal modes into transversal VCR modes [5]. Problem of energy transformation between longitudinal and transversal levels of freedom also is characteristic for theory of quark-gluon plasma investigated in collider experiments for colliding heavy ions [48] (see also references in [48]). In [48], it is paid special attention to the new point of view on the role of medium emitting coherent radiation of gluons in these experiments. For analysis of these experiments, it may be found important conclusions of the present work and also obtained in [49-52]. Actually, obtained above conclusion on permissibility of VCR+ mechanism by nuclear medium in the course of its interaction with $K^+$ and $K^0$ mesons, is found possible only in the frame of developed in [4, 5] and in the present work VCR and VCRA theory. For experiments LHC-ATLAS, CMS, the new VCR theory will too open additional possibility in observed data explanations which is not possible to explain on the base of usually applied VCR theory [3].

Obtained in [49-52] conclusion about relation of cosmological evolution to creation from vacuum of super light scalar bosons (mass of which is obtained as result of exact solution without singular modification of common relativity theory equations) allows to give new view also on the whole problem of seeking for Higgs boson [53].

Thus far, theory developed in [4, 5] and also in [49-53] and in the present work gives base for a new view on the vacuum physics as non-equilibrium medium and physics of virtual states of conventional medium. The latter as it is shown are able to be transformed for above threshold non-equilibrium conditions in real VCR and VCRA quanta.

We give also the new foundation for the developing of the quantum superfluid threshold velocity Landau theory modification, the new theory of super radiation in BEC and the new theory of the Compton effect in medium on the base of quantum VCR theory, proposed in [4,5] and in present paper.